\documentstyle[12pt,epsfig]{article}
\begin{document}
\title{Group--theoretical structure of quantum measurements and equivalence principle}
\author{ A. Camacho
\thanks{email: acamacho@aip.de} \\
Astrophysikalisches Institut Potsdam. \\
An der Sternwarte 16, D--14482 Potsdam, Germany.}

\date{}
\maketitle

\begin{abstract}
The transverse group associated to some continuous quantum measuring processes is analyzed
in the presence of nonvanishing gravitational fields. This is done
considering, as an example, the case of a particle whose coordinates are being monitored. Employing the so--called restricted path integral formalism,
it will be shown that the measuring
process could always contain information concerning the gravitational field.
In other words, it seems that with the pre\-se\-nce of a measuring process
the equivalence principle may, in some cases, break down.
The relation between the breakdown of the equivalence principle,
at quantum level, and the fact that the gravitational field could act always as a decoherence environment,
is also analyzed.
The phenomena of quantum beats of quantum optics will allow us to consider the possibility that the experimental corroboration
of the equivalence principle at quantum level could be taken as an indirect evidence
in favor of the quantization of the gravitational field, i.e., the quantum properties of this field
avoid the violation of the equivalence principle.

\end{abstract}
\newpage
\section{Introduction}
\bigskip

The detection of the effects of a gravitational field upon a quantum
system was ca\-rried out for the first time 25 years ago [1]. Since then,
this kind of experiments have achieved an impressive accuracy and feasibility [2, 3, 4], and
they involve the interaction of
gravity with an intrinsically quantum--mechanical quantity [1, 2], and therefore depend
upon Planck's constant $\hbar$ and Newton's universal gravitational constant $G$ [5]. This
last remark means also that the equivalence principle [6] can be analyzed at the quantum
realm.

In order to avert any misunderstanding here we follow
the definitions of [6], in other words, the so--called weak equivalence principle (WEP),
which is based on the principle that the ratio of the inertial mass to the gravitational--passive--mass is the same for all bodies,
asserts that {\it gravitation is locally equivalent to an acceleration of the reference frame}.
Einstein equivalence principle (EEP)
means here: {\it for every pointlike event
of spacetime, there exists a sufficiently small neighborhood such that in every
local, freely falling frame in that neighborhood, all the laws of physics obey
the laws of special relativity}.

The validity of the classical WEP has been verified with a high
precision [7] (for a recent description of the current experimental status in this context see [8]) and therefore it could be a natural assumption to
accept the validity of this principle
at quantum level.
Nevertheless, there are already results [9] which
prove that even the simplest classical kinematical concepts show conceptual difficulties when extra\-po\-lated into the quantum regime.
For instance, there is
no consistent definition of the time--of--flight probability density in the
case of a quantum particle, without classical counterpart, falling freely in
a gravitational field. The possibility of the emergence of a violation of EEP
in the case of a freely falling atom, which is being continuously monitored,
as a consequence of the coupling with the involved detecting device, has also been analyzed by Viola and Onofrio.
At this point it is noteworthy to comment that in this last work the problem has
not been addressed from the point of view of the symmetry properties of the measuring process.

Hence at this point we may pose the following question:
is the equivalence principle also valid at the quantum realm? Though there are works
that claim the validity of this principle in the quantum regime [10], we may already also find 
some results which single out quantum characteristics that could lead to the violation
of the equivalence principle [11, 12, 13].

Despite its spectacular success, quantum theory is beset with conceptual problems [14],
for instance, the problem of the classical limit of this theory, which is closely related
with the notions of ``observation'' and ``measurement'', comprises one of the more
long--standing conundrums in modern physics [15].

In this respect, one of the ideas that could solve this problem is the so--called
decoherence model (DM) [16], in which two systems (usually called collective system and environment, here we use the terminology appearing in [14], see page 268)
are coupled by an interaction Hamiltonian, such that it destroys quantum interferences at a macroscopic level.

The fact that gravity couples universally to all forms of matter [17] could lead to an
explanation of the emergence of a classical spacetime from a quantum model [18].
The possible role of the gravitational field, as a decoherence environment, has already been considered [19].
We may also find some approaches in which gravity plays a fundamental role in a 
possible modification of the formalism of quantum mechanics [20].

In the present work we will explore the following question: is there a relation between
the validity of EEP at quantum level and the role of gravity as a decoherence environment? This question has already been considered [21], but now it will be
done from a different perspective, i.e., analyzing the behaviour of the transverse group of a continuous measuring process.
In order to address this issue the group--theoretical properties of a particular quantum measurement
process in a curved manifold will be considered. It will be shown,
using the so--called restricted path integral formalism (RPIF), that the process that monitors
the coordinates of a quantum particle could always contain information concerning the gravitational field,
even when the corresponding  measuring process takes place, completely, in a locally flat reference frame.
In other words, it seems that with the presence of a measuring process
EEP breaks down. From this last fact it will be claimed that gravity could act always
as a decoherence environment, i.e., even local experiments (this phrase means those experiments
carried out within the corresponding tangent space) could suffer the action of the corresponding
gravitational field.

A possible connection of this last fact with the quantization of the gravitational field is also explored.
It will be claimed (resorting to an already known result from quantum optics) that this violation of EEP could disappear under the presence of
a quantum theory of gravity. In other words, if the involved model contains a classical gravitational field,
then EEP will be always violated. We may rephrase this stating that the quantum properties
of the field avoid the violation of EEP.
\bigskip

\section{Gravity and group--theoretical structure}
\bigskip

\subsection{Restricted path integral formalism}
\bigskip

The group--theoretical structure of quantum continuous measurements has been a\-nalyzed always in
the context of flat--manifolds, i.e., the possible effects of gravity
upon this aforementioned group--theoretical structure have not been considered [22].
A possible reason for this lies in the fact that the validity of EEP in connection
with quantum continuous measurements is always postulated. Nevertheless, there is already some evidence [9, 13]
which seems to indicate that this assumption could not always be valid.

Among the already existing attempts to solve the quantum measurement problem [14] we may find
the so--called restricted path integral formalism (RPIF) [23].
This formalism schematizes a continuous quantum measurement with the introduction of a restriction on
the integration domain of the corresponding path integral.
This last condition can also be reformulated in terms
of a weight functional that has to be considered in the path integral.

Let us suppose to have a particle exhibiting one--dimensional movement.
The amplitude $A(q'', q')$ for this particle to move from the point $q'$ to the point $q''$ is called propagator.
It is given by Feynman [24]

\begin{equation}
A(q'', q') = \int d\left[q\right]\exp\left({i\over \hbar}S_{[q]}\right),
\end{equation}

\noindent here we must integrate over all the possible trajectories $q(t)$, and $S_{[q]}$ is the action of the system, which is
defined as

\begin{equation}
S_{[q]} = \int_{t'}^{t''}dtL(q, \dot{q}).
\end{equation}

Let us now suppose that we continuously measure the position of this particle,
such that we obtain as measurement output a certain function $a(t)$. In other words, the measuring process gives the value $a(t)$
for the \-coor\-di\-na\-te $q(t)$ at each time $t$, and this output has associated a certain error $\Delta a$, which is determined by the
experimental resolution of the measuring device. The amplitude $A_{[a]}(q'', q')$ can be now thought of as a probability amplitude for the continuous measuring process to give the result $a(t)$.
Taking the square modulus of this amplitude allows us to find the probability density for different measurement outputs.

Clearly, the integration domain in the Feynman path--integral should be restricted to those trajectories that match with the experimental output.
RPIF says that this condition can be introduced by means of a weight functional $\omega_a[q]$ [23].
This means that expression (3) becomes now

\begin{equation}
A_a = \int d\left[q\right]\omega_{a}[q]\exp\left({i\over \hbar}S[q]\right).
\end{equation}
\bigskip

At this point it is noteworthy to comment that the action appearing in this last expression does not include the measuring device,
i.e., all the information about the experimental apparatus is contained in the weight functional.
The introduction of $\omega_{a}[q]$ does not imply the violation of the uncertainty relations.
Indeed, considering the case in which position is being monitored (see chapter 6 of [23]), it
is readliy seen that RPIF includes the unavoidable back reaction effect of the measuring device, which is a direct consequence of Heisenberg's uncertainty relation.

\subsection{Group--theoretical structure}
\bigskip

The group--theoretical properties of quantum continuous measurements comprises two different kinds
of structures [22], namely, the transverse group (which describes the homogeneity of the space
of measurement outputs, and therefore transforms alternative measurement outputs into each other),
and the longitudinal semigroup (it accounts for the evolution of the corresponding
quantum system subject to a continuous measurement).

In the present analysis we will focus on the transverse group. The idea behind
this is to find a violation of EEP using as core of our study a very simple, but illustrative situation,
i.e., the case of a particle whose coordinates are being monitored.

The fundamental assumption here is to accept that, before the evaluation of any
amplitude, all possible measurement outputs are equivalent. As pointed out in
[22], this condition means that there is some group $G$ (the transverse group)
which acts transitively on the set of all possible measurement outputs.

If $g\in G$, then the action upon measurement outputs, $\alpha$, and paths, $[q]$, are, respectively

\begin{equation}
g:\alpha \rightarrow g\alpha,
\end{equation}
\bigskip

\begin{equation}
g:[q] \rightarrow g[q].
\end{equation}
\bigskip

Clearly, the weight functional contains
all the information about the interaction between measuring device and measured system,
and in consequence any existing symmetries associated with the measuring process
will imply the invariance of $\omega_{a}[q]$ with respect to the group(s) related with the
corresponding symmetries.

Let us now suppose that the measuring device is embedded in a flat manifold, and
that $G$ is a symmetry of the corresponding experimental construction. In other
words, it is fulfilled that

\begin{equation}
\omega (\alpha, [q])= \omega (g\alpha, g[q]),
\end{equation}
\bigskip

\noindent here $g\in G$.

As a particular experiment let us assume that the measuring device monitors the coordinates
of our particle, and that the symmetry group, $G$, describes shifts in the configuration
space. Hence, $G$ is defined as [22]

\begin{equation}
G= {\Big\{}c(t)\vert ~t'\leq t\leq t''{\Big\}}.
\end{equation}
\bigskip

At this point it is noteworthy to mention that $G$ contains all possible (smooth) curves $c(t)$.

The action of $g\in G$ upon $\alpha$ and $[q]$ is

\begin{equation}
g\alpha= {\Big\{}q_{\alpha}(t) + c(t)\vert ~t'\leq t\leq t''{\Big\}},
\end{equation}
\bigskip

\begin{equation}
g[q]= {\Big\{}q(t) + c(t) \vert ~t'\leq t\leq t''  {\Big\}},
\end{equation}
\bigskip

\noindent where $q_{\alpha}(t)$ is the trajectory related to the measurement output $\alpha$.

Let us now suppose that the corresponding measuring device is located in a
region where a non--uniform gravitational field exists.
If $x\in M$, where $M$ denotes the involved curved manifold, then we may find a, small enough, neighborhood around
$x$ such that, according to EEP,  in this neighborhood the laws of physics are the laws of
special relativity [6].  Here we consider that the experiment takes place within this
neighborhood, i.e., the measuring device is located completely in this neighborhood.
At this point it is noteworthy to comment that in the case of a non--uniform gra\-vi\-tational field the involved neighborhood will not
cover the whole manifold. This very known fact will play a very important role in 
the possible breakdown of EEP in connection with the invariance properties of continuous quantum measurements.

In the case of a flat spacetime our group $G$ contains all possible (smooth) curves $c(t)$.
At this point we will focus on the following question: under the presence of a non--uniform gravitational field, which are the symmetries (if any)
of our measuring process? This may be rephrased saying: is there any group, $\tilde{G}$, under whose action
the corresponding weight functional is invariant (in the sense of expression (6))?
\bigskip

The answer to this question seems to have, at least, three possibilities:
\bigskip

1) $\tilde{G}$ contains all possible (smooth) curves, i.e., $\tilde{G} = G$ .
Clearly, if $x$ is a point in the corresponding manifold, $M$, then, as is already known [6, 17],
the validity region of the involved tangent space, $T_xM$, is in the case of a curved manifold 
limited, i.e., in a general gravitational field Lorentz frames are local, and not global. From the definition of $G$ (expression (7)), it is readily seen that we may find a
path $[q]$ and an element $g\in G$, such that $g[q]$ has some of its points outside
the validity region of $T_xM$.
In this situation, if $g[q]\not \in T_xM$, then in order to describe the invariance properties of our
ex\-pe\-ri\-ment (which takes place, completely, in $T_xM$) we must include the tangent spaces of some other
points in $M$ (because $T_xM$ does not cover the whole manifold). In other words, having $T_xM$ does not suffice to describe the
invariance properties of our measuring process, i.e., we must also have further information
concerning the geometry of the involved manifold. This fact implies that if we want to describe
the invariance pro\-per\-ties of a local experiment we do need information about the
gravitational field. This last assertion does not match with EEP. Indeed, this principle affirms that in
order to describe the measurement outputs of a local experiment only the laws of special relativity and
$T_xM$ are required.
\bigskip

2) $\tilde{G}$ does not include all possible (smooth) curves,  i.e., $\tilde{G} \not = G$, but only those
that lie completetly within $T_xM$. We may reformulate this last assertion stating that $G$ must contain information about the gravitational field.
Indeed, a nonvanishing gravitational field imposes restrictions upon $G$, such that
it becomes $\tilde{G}$. But in the case of a flat spacetime the invariance properties are described by $G$.
In other words, in this second alternative EEP breaks also down. Indeed, it is possible to detect
(by means of the invariance properties of the corresponding measuring process)
the presence of a gravitational field. This last fact also implies that the measuring device
is affected by the gravitational field, even when it lies completely within $T_xM$.
This may be rephrased stating that the  corresponding weight functional must depend upon the gravitational field,
even when the experiment is carried out within the validity region of our locally
flat coordinate system.
\bigskip

3) Under the presence of a nonvanishing gravitational field, $\omega(\alpha, [q])$ is not inva\-riant
under shifts in the configuration space. Hence once again the weight functional must depend upon the
gravitational field, and therefore in this case EEP breaks also down.
\bigskip

\section{Discussion}
\bigskip

\subsection{Gravity as decoherence environment and invalidity of EEP}
\bigskip

The previous arguments seem to indicate that in curved manifolds, there are
measuring processes whose invariance properties could render a violation of EEP.
The symmetry properties in a Minkowskian spacetime are not the same as those emerging in
a freely falling coordinate system inmersed in a curved manifold.

From cases (2) and (3) we could write the weight functional related to a measuring
process as

\begin{equation}
\omega = \omega (\sigma, \tilde{g};~\alpha, [q]),
\end{equation}
\bigskip

\noindent where the parameters $\sigma$ and $\tilde{g}$ point out, explicitly, the dependence of
the co\-rres\-ponding weight functional upon certain characteristics of the measuring
apparatus (for instance, its experimental resolution), and upon the gravitational field, respectively.

From the role that the involved gravitational field, $\tilde{g}$, plays in expression
(10), we could be tempted to assert that the gravitational field acts, always, as part
of a measuring device. In other words, the breakdown of EEP at quantum level would be closely related to the
fact that the gravitational field could act always as a decoherence environment,
and in consequence it would define a ``universal'' environment.
The possibility of a collapse of the wave function driven by gravitational effects has already been considered [19, 21, 25].
Nevertheless, the new ingredient put forward in the present
work states a connection between the in\-va\-li\-dity at quantum level
of EEP and the possibility of having gravity as a decoherence environment.
\bigskip

\subsection{Validity of EEP and quantum gravity}
\bigskip

It has already been claimed [26] that gravity--wave interferometers
could probe the fuzziness of space--time. This claim shows us that the detection of
effects coming from a quantum theory of gravity may not lie very far from the
present technological possibilities. Hence, at this point we wonder
if this conceptual inconsistency between general relativity and quantum measurement theory
renders some evidence for the existence of a quantum theory of gravity.

Starting with the case of the so--called quantum beat phenomena [27]
(and also resorting to the similarity that Maxwell's equations and Einstein linearized equations show [17]) we wonder if this violation of EEP
disappears with a quantized gravitational field.
We will see that this also implies that any experiment which, in the
future, could prove the validity at quantum level of EEP would be an
indirect proof of the existence of a quantum theory of gravity.

To this end let us remember that
in quantum optics a calculation which treats matter and radiation on the same footing (matter is described
quantum--mechanically and radiation along the ideas of quantum electrodynamics) explains the missing beats that the case
of a three--level atomic system of $\Lambda$ type shows [28], and that these calculations
are also compatible with quantum measurement theory.

These missing beats can not be
understood if light is described according to Maxwell's equations, even if they
are complemented with vacuum fluctuations. An explanation to this lies in the fact that these calculations contradict quantum measurement theory.

Summing up, in the case of quantum beats, if the field is treated classically, then the results do not match with 
quantum measurement theory, but the problem di\-sa\-ppears with the introduction of a quantized electromagnetic  field. In the present work we have a similar situation, 
matter has been treated quantum--mechanically, the involved field, gravity, is a purely 
classical entity, and additionally we have assumed the validity of quantum measurement theory.
Having in hindsight the previous arguments of quantum optics, it is no surprise at all
that a contradiction appears precisely at this point. Hence, we could re--interpret our results asserting that
the violation of EEP is a consequence of having a classical gravitational field, and not a quantum one.
In other words, if EEP is corroborated, experimentally, at quantum level, then
the question at this point would read: could this fact be an indirect evidence in favor of a quantum theory of gravity?
\bigskip

\subsection{Open topics}
\bigskip

As has been stated from the very begining (see section 2.2), we have restricted ourselves to
the case of the transverse group. Hence, an additional issue in this context comprises the analysis of the consequences
of the inclusion in the present model of the longitudinal semigroupoid,
and as it is already know [23], this semigroupoid (if the Hamiltonian does not depend explicitly
on time we have a semigroup) is related to the time evolution under continuous measurement of the corresponding system.
For instance, knowing that decoherence may be closely related to the emergence of an arrow of time [16] and
accepting that the violation of EEP could be connected with decoherence, it would be interesting to explore the possible interelations among
the gravitational field, the longitudinal semigroup, and the breakdown of the time--reversal symmetry.
This is another issue that has to be addressed.

Finally, a word must be said about the possible consequences of taking into account
the measurement process of a relativistic particle. It has already been proved [29] that
in the case of the position monitoring of a relativistic particle the resulting
amplitude is exponentially small outside the light--cone of the space--time point of the
corresponding measurement output. This last remark could modify in a substantial manner, for instance, our
symmetry group (expression (7)), and therefore also some of the corresponding conclusions.
This relativistic analysis is currently an open problem [30], and will be addressed in a separate paper.
\bigskip

\subsection{Conclusions}
\bigskip

We have shown that the transverse group associated to the continuous monitoring
of the position of a quantum particle, inmersed in a gravitational field, contains information concerning this field.
In other words, it seems that with the presence of a measuring process
the equivalence principle may, in some cases, break down.
The possible relation between the breakdown of the equivalence principle,
at quantum level, and the fact that the gravitational field could act always
as a decoherence environment has also been analyzed.
The phenomena of quantum beats of quantum optics allowed us to consider the possibility that the experimental corroboration
of the equivalence principle at quantum level could be taken as an indirect evidence
in favor of the quantization of the gravitational field, i.e., the quantum properties of this field
avoid the violation of the equivalence principle.

\bigskip
\bigskip
\bigskip

\Large{\bf Acknowledgments.}\normalsize
\bigskip

The author would like to thank A. A. Cuevas--Sosa for his 
help, and D.-E. Liebscher for the fruitful discussions on the subject. 
The hospitality of the As\-tro\-phy\-si\-ka\-li\-sches Institut Potsdam is also kindly acknowledged. 
This work was supported by CONACYT (M\'exico) Posdoctoral Grant No. 983023.

\end{document}